# A time for monsters: Organizational knowing after LLMs

Samer Faraj

Joel Perez Torrents

Saku Mantere

Anand Bhardwaj

Desautels Faculty of Management
McGill University

**Abstract**

Large Language Models (LLMs) are reshaping organizational knowing by unsettling the epistemological foundations of representational and practice-based perspectives. We conceptualize LLMs as Haraway-ian monsters, that is, hybrid, boundary-crossing entities that destabilize established categories while opening new possibilities for inquiry. Focusing on analogizing as a fundamental driver of knowledge, we examine how LLMs generate connections through large-scale statistical inference. Analyzing their operation across the dimensions of surface/deep analogies and near/far domains, we highlight both their capacity to expand organizational knowing and the epistemic risks they introduce. Building on this, we identify three challenges of living with such epistemic monsters: the transformation of inquiry, the growing need for dialogical vetting, and the redistribution of agency. By foregrounding the entangled dynamics of knowing-with-LLMs, the paper extends organizational theory beyond human-centered epistemologies and invites renewed attention to how knowledge is created, validated, and acted upon in the age of intelligent technologies.

# Introduction

The emergence of intelligent technologies, capable of synthesizing knowledge autonomously from human supervision (Acharya et al., 2025; Bender et al., 2021), in particular those integrating Large Language Models (LLMs), is reshaping how knowledge work is conducted in organizations. Scholars in strategy and organization have examined how knowledge is created, shared, and retained, offering theories anchored in two dominant traditions. A representational perspective treats knowledge as an object that can be codified, stored, and transferred across contexts (Alavi and Leidner, 2001; Argote et al., 2003; Nonaka and Takeuchi, 1995). A practice-based perspective views knowing as embodied and situated, enacted through participation in specific practices (Hadjimichael and Tsoukas, 2019; Nicolini, 2011; Orlikowski, 2002). While these approaches have yielded rich insights, both rest on a critical assumption: that intelligence and knowledge production reside solely in humans, with technologies serving only as containers or conduits.

The introduction of LLMs[1] challenges this assumption. Unlike earlier technologies designed to store or transmit knowledge, LLMs exhibit agential qualities in interaction. Going beyond the simple retrieval of codified information, they generate new knowledge by identifying statistical patterns and surfacing analogical relations across vast corpora. This capacity positions them as active participants in knowledge work: reframing problems, proposing solutions, and revealing previously hidden connections. In doing so, they transform how organizational actors generate new knowledge. For organizations, this increased entanglement between human agents and artificial ones in knowledge work represents a major challenge.

[1] We focus on Large Language Models (LLMs) as foundational building blocks for more complex intelligent systems, including agentic and multimodal AI, that reshape organizational knowing. While LLMs are increasingly applied beyond text to sound, images, and video, our analysis here is limited to text-based models.

From organizational capabilities to task and job distribution, to roles and networks, and even the boundary of the firm, many of the conceptual bedrocks of our field were developed for a world in which intelligence was unambiguously human. Scholars have begun to raise important questions about how AI, and large language models (LLMs) in particular, may alter existing modes of knowledge production and organizational knowing (e.g., Argote et al., 2022; Bailey et al., 2022; Cornelissen et al., 2024; Grimes et al., 2023). Yet much of this work remains exploratory, often framed as calls for new theorizing rather than offering settled answers. Existing theories of knowledge either assume codification into human-understandable categories or embed knowing in human bodies and practices. Neither of these epistemic foundations can account for how the kind of knowledge work LLMs perform on vast textual corpora or for how they interact with organizational knowing processes. Our theoretical vocabulary must adapt and address questions like: how do LLMs shape the production of organizational knowledge? How does this novel form of knowledge production differ from previous ones? Consequently, how should organizational actors engage with these technologies?

We begin by reviewing what we see as two principal approaches to knowing in organizations (representational and practice-based). We find that at the core of each of them lies analogizing: the process through which organizations generate, validate, and expand knowledge by identifying and exploiting structural similarities between known and unknown domains of meaning (e.g., Cornelissen et al., 2011; Hofstadter and Sander, 2013; Ketokivi et al., 2017; Tsoukas, 1993).

After this review, we build a third view of knowing that is sensitive to foundational differences in how humans and LLMs reason with analogies. To make sense of this emerging human–AI entanglement, we draw on posthumanist philosopher Donna Haraway's (1992, 2016) concept of the monster, not as a creature of fear or fantasy, but as a boundary-crossing figure that challenges established distinctions such as human/machine, nature/culture, and organism/technology. Accordingly, we position LLMs as partial, unpredictable, and generative contributors to organizational knowing, shifting our attention from what tasks LLMs can do to what kinds of entanglements they enact. By framing LLMs in this way, we outline the contours of a third epistemological stance, knowing-with-LLMs, that acknowledges the emergent human–AI entanglements and invites organizational theorists to rethink how knowledge is created, evaluated, and acted upon in the age of intelligent technologies.

## Organizational knowing and the challenge of LLMs

The capacity to know collectively, to act knowledgeably at the collective level and to incorporate knowledge are central tenets to what we understand as an organization. Adam Smith's (1776) classical example of a pin factory demonstrates why organizations such as firms exist: coordinated action between several workers in technological arrangements is a far superior means of producing iron pins than a single artisan completing each step alone. In the parlance of contemporary organizational theory, the boundary of the pin factory is drawn by the organizational knowing of pin-making (Spender and Grant, 1996). Organizations can be understood as composites of distributed knowledge (Tsoukas, 1996), integrated into coordinated ways of knowing that facilitate the solving of key organizational problems (Zahra et al., 2020).

In such problem-solving, analogizing, *i.e.* the process of reasoning by analogy, underpins how actors make sense of ambiguity, navigate novelty, and construct shared meaning (Ketokivi et al., 2017). Analogizing supports both explanatory and generative work, enabling individuals and groups to map the unfamiliar onto the familiar, reframe problems, and craft narratives that guide action (Cornelissen et al., 2011; Tsoukas, 1993). As Hofstadter and Sander (2013: 3) put it, analogies are the "fuel and fire of thinking," not a decorative feature of language but the very engine of cognition. Analogies operate through metaphors, prototypes, and stories, shaping strategy, learning, and innovation (Cornelissen and Durand, 2014). They tie the past to the present, resituating old or similar experiences and thus allowing actors to cope with new situations via cognitive and narrative templates for action.

Analogizing fosters both innovation and organizational learning, providing an engine for expanding domains of knowledge. In innovation studies, developing solutions involves defining the structure of the problem followed by a search for relevant analogue solutions that can then be customized into a specific solution (Herstatt and Kalogerakis, 2005). In strategic management, analogizing informs strategic framing, business model innovation, and opportunity recognition, particularly under conditions of uncertainty (Lovallo et al., 2012). More generally, many organizational theories are fundamentally analogical in nature (*e.g.*, organization as an organism), allowing actors to frame organizational phenomena into terms that are more foundational and insightful (Ketokivi et al., 2017; Cornelissen et al., 2011). Generating analogies renders abstract and unfamiliar concepts intelligible and actionable, particularly in interdisciplinary collaborations and emerging fields where shared vocabulary is limited.

The process of analogizing plays an essential role in the two commonly acknowledged foundational epistemologies for organizational knowing (Cook and Brown, 1999): representationalism, which treats knowledge as an object, and a practice-based perspective, which understands knowing as an embodied and situated activity. As detailed below, each epistemology carries implicit assumptions about the nature and role of technology in organizational knowing. We argue, however, that such assumptions may no longer hold in the context of intelligent technologies such as LLMs.

### *Representationalism-based knowing*

Representationalism conceptualizes knowledge as a resource that can be codified, stored, and transferred to generate competitive advantage (Argote et al., 2003; Grant, 1996; Nonaka and Takeuchi, 1995). From this perspective, technology functions as a passive container and conduit for codified expertise: documents, databases, and systems that archive best practices and lessons learned (Alavi and Leidner, 2001). The aim is to convert tacit human knowledge into explicit forms that can be efficiently retrieved and disseminated. Analogies are important ways to represent and clarify concepts. They are codifiable as structured heuristics or templates that can be stored, retrieved, and reused across contexts. They function as cognitive assets that enhance efficiency and reduce cognitive load. Their value lies in transferability and repeatability: the ability to be archived, indexed, and reused in new situations

LLMs challenge this passive conception of technology. Rather than simply retrieving codified content, they generate context-sensitive responses by synthesizing across vast corpora, effectively dissolving the boundaries between storage, retrieval, and inference. In doing so, LLMs also appear to ease persistent challenges in transferring knowledge across technical, functional, and organizational boundaries, difficulties often attributed to the stickiness of tacit knowledge and the translation across divergent interpretive frameworks (Carlile, 2004; Nonaka, 1994; Szulanski, 2003).

Rather than simply expand organizational knowledge by augmenting human capacity or replacing it altogether through automation (Raisch and Krakowski, 2021), LLMs introduce a generative capacity that challenges foundational assumptions in organizational knowledge theories. Most notably, they disrupt the representationalist epistemology that underpins models such as Nonaka's

SECI framework, which presumes a human-centered cycle of tacit-to-explicit knowledge conversion grounded in embodied practice. Because their statistical processes dissolve the distinction between storage, retrieval, and inference, LLMs blur the previously foundational distinction between tacit and explicit forms of knowledge. Useful and even insightful outputs are produced via statistical "inference" with minimal human understanding of the process. Conceptually, if knowledge is dynamically generated in interaction with a probabilistic model, then core concepts like "knowledge transfer," "best practices," or even "expertise" need rethinking. We propose that these systems with their black-boxed performance reconfigure how knowledge is enacted, circulated, and legitimated in organizations (Faraj et al., 2018; Faraj and Leonardi, 2022; Hinds and von Krogh, 2024; Kellogg et al., 2020). This reconfiguration breaks foundational assumptions at the heart of the knowledge-based view (see Grant, 1996), such as the strategic significance of a firm's specific knowledge, the role of the firm in integrating the specialized knowledge residing in individuals, and the importance of structures and rules in supporting knowledge integration.

## *Practice-based knowing*

The practice-based perspective views knowing as enacted in embodied, situated, and material engagement with the world (Hadjimichael and Tsoukas, 2019; Nicolini, 2011; Orlikowski, 2002). In this perspective, knowing is not a disembodied resource or a cognitive possession, but instead, emerges from situated engagement with the world through ongoing action, interaction, and embodied practice. Technologies are not neutral tools; their epistemic contribution emerges through use, as actors enact patterns of engagement that shape both practice and meaning (Anthony, 2021; Jussupow et al., 2021; Lebovitz et al., 2022). As Orlikowski (2002:249) puts it, "knowing is not a static, embedded capability or stable disposition of actors, but rather an ongoing social accomplishment." In this epistemology, analogizing is a situated activity: a way for actors to make sense of unfolding events and compare the present challenges to past or related situations. It is an active process maintained through repeated participation in practice, and shaped by habits, tools, norms, and local routines.

Here, too, LLMs pose a challenge. The assumption that humans remain the primary locus of knowing is unsettled when systems generate summaries, diagnoses, or reframings that mimic expert reasoning without embodied grounding. This emergent, seemingly alien, expertise complicates the practice-based view: outputs grounded in statistical inference now intervene directly in knowledge work, performing epistemic functions once reserved for humans (Anthony et al., 2023; Scarbrough et al., 2024). As Sergeeva et al. (2023) note, we are entering an era where technologies are actively transforming the foundations of knowledge work. As a result, LLMs are emerging as a novel kind of epistemic contribution despite lacking the social, situated, and embodied basis of human knowing that this perspective supposes.

For the practice-based perspective, acute tensions are emerging. If tacit knowledge acquisition relies on bodily indwelling and focal–subsidiary awareness (Hadjimichael et al., 2024; Polanyi, 1966), then the epistemic involvement of LLMs risks weakening the traditional learning arc. This may introduce new pathways to expertise that bypass the Polanyian "know-how" typically developed through iterative, embodied engagement with material work. While these developments may not render practice-based knowing obsolete, they nonetheless challenge foundational assumptions about how knowing emerges through situated engagement, socialized participation, and embodied learning.

## *Toward knowing-with-LLMs*

Taken together, these challenges expose the limits of both epistemic traditions. In that sense, LLMs are Haraway-ian monsters, disrupting the assumptions of each and blurring classical epistemological divides between representationalism and practice-based views. Technology can no longer be treated as external or passive. Increasingly entangled with human actors, it is not merely executing predefined activities; it is redefining epistemic tasks themselves, such as how knowledge is accessed, interpreted, validated, and communicated, transforming what it means to “know” in an organization.

This disconnect between human expertise and algorithmic output creates novel challenges for knowledge work in organizations. It has given rise to new interrogation practices that attempt to explain the source of differences (Lebovitz et al., 2022; Jussupow et al., 2021), or for the deployment of knowledge brokers who translate algorithmic results into domain-relevant insights (Waardenburg et al., 2022). More broadly, where human expertise once represented the primary bottleneck in organizational knowing, LLMs now generate a surplus of information, intensifying the need for new strategies of validation and sensemaking (Grimes et al., 2023; Hannigan et al., 2024).

We propose a relational perspective of knowing-with-LLMs that examines how human–machine entanglements reorganize the epistemic foundations of organizational knowing. This perspective shifts the question from what a technology’s effects are, to how this entanglement reshapes the social and epistemic relations that constitute organizational life (Bailey et al., 2022; Faraj and Leonardi, 2022; Hinds and von Krogh, 2024). Rather than centering on the technology itself, it foregrounds how constellations of humans and LLMs reconfigure data flows, interpretive practices, and relations of expertise. In this light, LLMs function not merely as tools or partners but as epistemic monsters in the Haraway-ian sense: hybrid actors that transgress established boundaries between tacit and explicit knowledge, between user and system, between knowing and unknowing. To explore these dynamics, we focus on an underappreciated but fundamental driver of knowledge and knowing: analogizing. By centering analogical reasoning, we show how LLMs co-participate in the production of knowledge and how organizations might learn to live with these new epistemic monsters.

## LLMs as analogy engines

The rise of LLMs introduces a novel mechanism for generating analogies, one that differs fundamentally from both traditional symbolic AI and embodied human cognition. Earlier AI systems working on analogies relied on explicit rules or limited feature-matching (Gentner, 1983; Holyoak and Thagard, 1995). Classic models for analogy-making such as Gentner's (1983) Structure-Mapping Theory framed analogy as a process of matching patterns of relationships across domains, often privileging deep causal correspondences over surface traits. By contrast, LLMs generate analogies through statistical pattern analysis across vast corpora, encompassing books, articles, blogs, and forum discussions.

At the core of LLMs' generative power is the way they represent words and textual units not as symbols with fixed meanings, but as vectors in high-dimensional space, based on statistical co-occurrence across vast text corpora. For example, "cat" as a vector (*e.g.,* [0.22233, 0.2334211, 0.4432, …]) is mapped based on its co-occurrence with other words. "Dog" will occupy a region near “cat” in the embedding space, reflecting the many contexts in which these two concepts appear (Mikolov et al., 2013). LLMs are also set apart from older neural network architectures by the “attention mechanism” (Vaswani et al., 2017) that weights the importance of different words in a sequence and enables the model to capture distant similarities and contextual nuance more effectively.

Analogical reasoning in LLMs thus arises from vector operations that map relationships from one region of the embedding space to another. A classic example would be the finding of “man is to woman what king is to …”. The answer is “queen”, and this analogical reasoning is derived from a vector operation in the embedded space: vector(“king”) - vector(“man”) + vector(“woman”) ≈ vector(“queen”). Likewise, the relationship between “Paris” and “France” is statistically analogous to “Berlin” and “Germany” because, in the training corpora, each capital is similarly positioned relative to its country. These derivations extend to more abstract domains, like recognizing parallel structures between the concepts of “organization” and “organizing” or “affordance” and “affordancing.” An LLM can even propose a newly minted concept, like “performativification,” by inferring that if “sign” leads to “signification,” then “performative” might lead to “performativification.”

This form of reasoning is fundamentally distinct from human analogizing which relies on conceptual isomorphism. Human inference draws on situated action, embodied experience, intuition, causal thinking, and meaning construction (Tsoukas, 1993, 2009; Hofstadter and Sander, 2013; Ketokivi et al., 2017; Yanow and Tsoukas, 2009). As Fodor (2000) argued in his critique of early computationalism, the human mind’s inferential capabilities depend on background knowledge, intentionality, and a tacit grasp of context. By contrast, LLMs generate analogies purely from statistical co-occurrence across immense corpora. When an LLM is tasked to explain how it generated a specific analogy, the machine will generate a human-friendly explanation that mimics a causal argument but masks the underlying statistical operations that were performed to reach that result.

In short, contrary to human analogizing, which is grounded in embodied or social knowing, LLMs function as probabilistic analogy engines. This shift has implications for both representational and practice-based epistemologies. From a representationalist view, analogies have long been treated as reusable cognitive templates, that is, heuristics that can be codified, stored, and applied across contexts to enhance decision-making. While LLMs appear capable to generate analogies on demand, their statistical derivation lacks the semantic stability or intentional structure assumed by the representational approach. From a practice-based view, analogizing is an embodied, situated performance shaped by lived experience and social context. Here, too, LLMs unsettle foundational assumptions by producing analogical insights without participating in practice or engaging with the world. In both cases, probabilistic generation challenges what analogies are, how they function, and what they mean in organizational knowledge work. In the next section, we explore how LLMs both extend and destabilize established modes of organizational knowing.

# Modes of analogizing with LLMs

The growing integration of LLMs raises important questions about how they increasingly participate in knowledge work by generating analogies that bridge unfamiliar domains, reframe problems, and offer novel comparisons. Analogies are essential for organizing complexity, making sense of uncertainty, and guiding action. Understanding how LLMs generate analogies offers a concrete entry point into examining their broader epistemic role in organizations. In particular, analyzing different forms of analogizing, especially along the dimensions of depth and distance, clarifies how LLMs reshape organizational knowing.

Analogies vary in their form and epistemic strength, ranging from loose metaphorical expressions that evoke imagery or impressions, to systematic mappings that align relational and functional structures across domains. Research from both cognitive science and organization theory suggests that analogizing varies along two key dimensions: depth and distance (Gentner, 1983; Hofstadter and Sander, 2013). *Depth* refers to the degree of structural alignment between source and target domains. *Surface analogies* rest on surface-level resemblance (e.g., visual or linguistic similarity), while *deep analogies* rely on shared relational or causal structures. This distinction also clarifies how metaphors differ from analogies: metaphors work at the level of language and discourse with limited structural isomorphism, while analogies systematically connect knowledge structures across domains. For instance, describing an organization 'as a plate of spaghetti' (Foss, 2003) is a vivid explanatory metaphor, but it does not enable predictions about organizational behavior. In contrast, stating that an airport is like 'a water distribution network' offers a deeper analogy: the bottlenecks in the system of pipes can be used to model check-in gates and security, limiting the "flow" of passengers (Ketokivi et al., 2017).

*Distance* captures the conceptual reach of analogizing. *Near analogies* operate across adjacent or familiar domains, while *far analogies* bridge distant or unrelated ones. The spaghetti metaphor is a distant surface comparison, as food and organizational structure are not normally linked. By contrast, analogies drawn from personhood, institutions, or biological organisms are closer to organizational theory because they have become conventional frames (Boxenbaum and Rouleau, 2011; Ketokivi et al., 2017). Far analogies, while riskier, can be highly generative since their interpretive potential remains unexplored. Crossing the dimensions of depth and distance produces four distinct forms of analogizing, summarized in Table 1.

| | **Near domain** | **Far domain** |
|---|---|---|
| **Surface analogies** | **Metaphoric exploitation**<br><br>LLMs retrieve familiar idioms and easy to grasp metaphors, enabling quick comprehension and alignment<br><br>*Risks*: reinforces clichés, blind spots, and superficial thinking | **Metaphoric exploration**<br><br>LLMs draw on shallow similarities to generate novel juxtapositions that spark creativity and break entrenched frames<br><br>*Risks*: misleading or unstable results; shallow engagement with underlying processes |
| **Deep analogies** | **Analogical exploitation**<br><br>LLMs surface structural parallels across adjacent domains, supporting incremental learning and transfer of heuristics<br><br>*Risks*: requires human expertise to verify accuracy and validate contextual fit | **Analogical exploration**<br><br>LLMs uncover structural patterns across distant domains, enabling reframing and paradigm-shifting insights<br><br>*Risks*: prone to hallucination, requires rigorous vetting |

**Table 1: Forms of Analogizing with LLMs**

Below, we explore how LLM-enabled analogizing reshapes organizational knowing based on the analytical framework of Table 1.

### *Metaphoric exploitation (surface + near)*

Metaphoric exploitation draws on well-established comparisons that emphasize visible features across familiar domains. These analogies facilitate rapid sensemaking and alignment around shared schemas, as in: “a KPI dashboard is like a car’s instrument panel” or “cutting meeting times is the low-hanging fruit of productivity.” These analogies resonate immediately with organizational members, assisting quick comprehension, facilitating communication, and reinforcing shared organizational culture and narratives. While easy to produce with LLMs, such analogies often lack novelty and may simply reproduce trivial comparisons that align with pre-existing frames. Their very familiarity can also perpetuate clichés, blind spots, and superficial thinking. Their value lies in speed and familiarity, but their epistemic depth is shallow.

### *Metaphoric exploration (surface + far)*

Metaphoric exploration links resemblant features across conceptually distant domains. These surface analogies may appear eye-catching or even important for they link an existing field to an unexpected one. It may bring about a new comprehension of a situation or offer a novel explanatory power to phenomena in one’s field by producing evocative juxtapositions, *e.g.*, strategy-making is like “gardening in shifting soil.” LLMs are particularly effective at generating shallow, linguistically driven associations across distant knowledge domains. While such metaphors can stimulate creativity and open new perspectives, they often lack depth or actionable clarity. Their value depends on critical interpretation and contextual validation by humans. Without this, surface-far analogies risk being misleading, overly simplistic, or unstable. Their organizational value depends on the reflective scrutiny applied by human users.

### *Analogical exploitation (deep + near)*

Analogical exploitation synthesizes structural parallels across closely related or familiar domains. Examples include transferring business models across adjacent industries to another or adapting pricing strategies across similar markets. Deep-near analogies rely on mapping causal mechanisms and structural patterns across near domains. These analogies offer actionable insights grounded in shared context and support problem-solving by exploiting deep relational similarities. Humans generate such insights through experience and reflective engagement, but LLMs can facilitate this process by surfacing statistical regularities in corpora. The difference lies in how structural similarity is detected: human inference is intentional and situated, whereas LLM output must be interpreted and validated. Without domain expertise, users may mistake statistical associations for causal explanations, leaving the relevance and accuracy of LLM-generated analogies uncertain.

### *Analogical exploration (deep + far)*

Analogical exploration seeks structural correspondences across distant domains to reframe understanding from one to another. Such cross-domain mappings can lead to paradigm shifts, changing how a phenomenon is conceptualized or addressed. A classic example is borrowing from immunology research to rethink IT security, creating novel defense strategies based on how the body adapts responses and self-regulates. Prior research highlights this quadrant as especially generative for innovation (Herstatt and Kalogerakis, 2005; Martins et al., 2015). In this quadrant, LLMs are particularly promising, however the results could be difficult to evaluate. Humans often lack the capacity to assess whether the analogies are genuinely relevant or if they are grounded in meaningful correspondences. Yet this quadrant carries the greatest epistemic risk: hallucinated patterns, inappropriate mappings, or seductive nonsense. Its promise lies in expanding the horizon of possible sensemaking; its danger lies in ungrounded analogies mistaken for insight.

Taken together, the four quadrants show how LLMs generate analogies across a spectrum from safe metaphors to radical reframing. The analysis of each quadrant highlights both the promise and the peril of knowing-with-LLMs: they expand the horizons of organizational sensemaking but also produce outputs that are uncertain and hard to evaluate. In the next section we turn to three essential challenges organizations must attend to as they learn to live with these epistemic monsters.

## Living with monsters: Inquiry, vetting, and agency challenges

We have argued that LLMs are better understood as analogy engines, capable of generating novel knowledge through large-scale statistical inference. They do not simply extend human capacities for knowing; they transform the very conditions that make knowing possible, reshaping the epistemic arrangements, social dependencies, and material infrastructures through which knowledge is produced. To conceptualize this transformation, we draw on the figuration of the monster: a boundary-crossing entity that unsettles epistemic order and long held assumptions. Monsters provoke fear. From the Latin *monstrum*, a gerund of "to remind, warn, instruct, or foretell", they became culturally entangled with something defying nature. Whether it is centaurs, werewolves, vampires, Frankenstein's creature, or stillborn malformed children labeled "monstrous births" by scientists, they are "inappropriate/d others": entities that do not fit into neat taxonomies and that challenge the logic of classification itself (Haraway, 1992). Today, LLMs threaten because of their status as 'unknown' quantities that defy classification and carry an undetermined potential to disrupt the status quo of power, knowledge, and identity.

Yet monsters are not only figures of fear but also of possibility. They can mark moments when familiar order, stable boundaries, and long-held assumptions begin to unravel and promising new forms of thinking, organizing, and becoming emerge. Haraway (2016), for example, introduced the cyborg as a monster-like figure precisely to challenge rigid biological and social categories and to propose an emancipatory feminist reframing that embraces hybridity and situated knowledge. Ray (2022) showed how the very foundations of human physiological understanding emerged from engaging with "monstrous" anomalies, revealing how deviance can become a source of epistemic progress. More broadly, STS scholars (*e.g.*, Jasanoff, 2005; Law, 1991) have shown how the prospect of producing monsters through genetic and biotech experimentation has continuously forced reappraisal and refinement of dominant ontologies and epistemologies. By revealing the constructedness and fragility of our systems of knowing, monsters invite us to experiment with potentially transformative assemblages of humans, technologies, discourses, and materials.

Thinking about LLMs as monsters exposes the fracture lines in organizational knowing. They are epistemic actors that compel us to reconsider who or what can inquire. As emergent epistemic actors, they compel us to reconsider what it means to inquire, and who or what is entitled to do so. They straddle the line between representation and practice, yet fit neither, generating knowledge without intention or subjectivity. Their monstrous status stems from ontological ambiguity: too mechanical to be recognized as subjects, too generative to be dismissed as mere tools, too entangled to remain external. By speaking without understanding or embodiment, they unsettle foundational assumptions about how knowledge is produced and by whom. As STS scholars remind us, categories are never neutral; they enact boundaries of inclusion and exclusion that favour some forms of knowing while excluding others (Bowker and Star, 1999). In line with Haraway, we can welcome these monsters as provocative companions that invite reflexivity, challenge settled notions of expertise and agency, and connect us to novel modes of inquiry. To grasp these implications, we highlight three dimensions that mark how LLMs are fundamentally reshaping organizational knowing.

First, LLMs transform the *nature of inquiry*, the collective processes of questioning, exploring, and sensemaking through which organizations generate and attend to knowledge. They do so by widening the scope of inference beyond what any individual expert or research community could encompass. Traditional inquiry has long been constrained by the limits of individual expertise, with most knowers developing depth within a single field and facing difficulty in traversing to other domains (Leone et al., 2021). LLMs unsettle this model by producing outputs that are plausible without being verifiable in reference to a ground truth. This undermines the authority of codified knowledge and raises risks of spurious precision or information overload. Because LLM responses vary based on user interaction history and query formulation, they can also contribute to knowledge drift and heightened equivocality. For those viewing knowing as embodied, situated, and enacted through participation in social practices, LLMs reconfigure knowing as well. As organizational members increasingly engage in sensemaking via these nonhuman actors, concerns arise about the erosion of tacit knowledge and the devaluation of the slow cultivation of tacit skills and experiential indwelling. In both traditions, LLMs overflow their conceptual boundaries: destabilizing repositories of codified knowledge on one hand and threatening the embodied ground of practical knowing on the other. In ways that are still emerging, LLMs enlarge the horizon of inquiry, opening up novel connections that, upon closer scrutiny, may turn out to be superficial and lack grounding (see Cornelissen et al., 2024; Hannigan et al., 2024). Thus, the emerging challenge surrounding inquiry is in establishing when and how machine-generated insights enhance inquiry and when they lead it astray.

Second, the epistemic promise of LLMs heightens the need for what we call *dialogical vetting*, a recursive, socially situated process for evaluating the truthfulness, relevance, and contextual fit of machine-generated outputs. Because their probabilistically generated output is not rooted in validated truths, their answers cannot be accepted at face value. Instead, they must be engaged dialogically: interpreted, interrogated, and repeatedly tested against situated knowledge. Dialogical vetting is not reducible to one-off verification but requires the engagement in a recursive practice of questioning, evaluating, reframing, and re-prompting, in which organizational knowers supply the contextual judgment that machines lack. Prompting itself becomes central as part of a new literacy necessary to engage with AI output (Mollick, 2024). An emphasis on dialogical vetting builds on and extends the organizational learning tradition, which has long emphasized the importance of shared validation practices (Argote et al., 2003). It aligns closely with practice-based approaches to knowing, with vetting becoming part of the unfolding inquiry process, grounded in collective participation and reflexive engagement (Hadjimichael and Tsoukas, 2019). Field studies of AI in use already point to the need for deeper validation practices (Anthony, 2021), interrogation of model outputs (Lebovitz et al., 2022), and two-way monitoring (Jussupow et al., 2021). Dialogical vetting thus becomes a collective discipline of discernment, through which organizational members continually anchor machine outputs in lived contexts. Without such a recursive engagement, inquiry risks being swamped by plausible but ungrounded associations; with it, the outputs of knowing-with-LLMs can trigger deeper reflection and more resilient forms of knowing.

Finally, living with these epistemic monsters requires a *reconsideration of agency*. If knowledge is no longer confined to human understanding or situated practice but emerges from ongoing interactions between humans and generative models, then questions of authorship, accountability, and authority must be rethought. Who, or what, can be said to "know" when an LLM reframes a strategic problem? Who bears responsibility when organizational action is guided by a knowing-with-LLM analysis? LLMs unsettle the distribution of epistemic agency within organizations, blurring the lines between tool and collaborator, between human judgment and machine output. Rather than remaining backgrounded tools, they redistribute human agency, pressing practitioners to assume new roles as curators, interpreters, and co-authors of machine-generated knowing outputs. This transformation involves more than human augmentation (Raisch and Krakowski, 2021) or the addition of a collaborative partner (Mollick, 2024). It points instead to a redistribution of agency, where knowing is enacted across assemblages of humans and machines (Leonardi, 2012; Orlikowski and Scott, 2008). In this entanglement, agency is never fully assigned as it emerges from, and shifts with, the dynamics of interaction and interpretation. Most importantly, these new modes blur lines of accountability: who can legitimately claim authorship, who bears responsibility for error, and how expertise, authorship, and accountability are established.

In this paper, we have argued that LLMs are best understood as epistemic monsters that destabilize established ways of organizational knowing while opening new possibilities. They transform inquiry by extending the scope of inference beyond human or disciplinary limits. They make dialogical vetting central, as recursive practices of questioning, reframing, and re-prompting become essential. Finally, they occasion a redistribution of agency as the lines between tool and human give way to entanglement. Like Haraway's monsters, LLMs are unsettling because they transgress boundaries, but they are also generative, compelling us to imagine alternative directions for organizational inquiry. To live with these monsters is to recognize both their risks and their promise. We invite organizational scholars to a deeper exploration of the new epistemic possibilities that knowing-with-LLMs are making possible.